\documentclass[prl,twocolumn,preprintnumbers,superscriptaddress]{revtex4-1}
\usepackage{amssymb}
\usepackage{amsmath}
\usepackage{amsfonts}
\usepackage{graphicx}
\usepackage{dcolumn}
\usepackage{bm}
\usepackage{verbatim}
\usepackage{color}
\usepackage{wasysym}
\usepackage{xcolor}
\usepackage{ulem}

\begin{document}

\title{Bose-Einstein condensates of microwave-shielded polar molecules}
\date{\today }
\author{Wei-Jian Jin}
\affiliation{Institute of Theoretical Physics, Chinese Academy of Sciences, Beijing 100190, China}
\affiliation{School of Physical Sciences, University of Chinese Academy of Sciences, Beijing 100049, China}

\author{Fulin Deng}
\affiliation{School of Physics and Technology, Wuhan University, Wuhan, Hubei 430072, China}
\affiliation{CAS Key Laboratory of Theoretical Physics, Institute of Theoretical Physics, Chinese Academy of Sciences, Beijing 100190, China}

\author{Su Yi}
\email{syi@itp.ac.cn}
\affiliation{Institute of Theoretical Physics, Chinese Academy of Sciences, Beijing 100190, China}
\affiliation{School of Physical Sciences, University of Chinese Academy of Sciences, Beijing 100049, China}
\affiliation{Peng Huanwu Collaborative Center for Research and Education, Beihang University, Beijing 100191, China}

\author{Tao Shi}
\email{tshi@itp.ac.cn}
\affiliation{Institute of Theoretical Physics, Chinese Academy of Sciences, Beijing 100190, China}
\affiliation{School of Physical Sciences, University of Chinese Academy of Sciences, Beijing 100049, China}

\begin{abstract}
We investigate the ground-state properties of the ultracold gases of bosonic microwave-shielded polar molecules. To account for the large shielding core of the inter-molecular potential, we adopt a variational ansatz incorporating the Jastrow correlation factor. We show that the system is always stable and supports a self-bound gas phase and an expanding gas phase. We also calculate the condensate fraction which is significantly reduced when the size of the shielding core of the two-body potential becomes comparable to the inter-molecular distance. Our studies distinguish
the molecular condensates from the atomic ones and invalidate the application of the Gross-Pitaevskii equation to the microwave-shielded molecular gases. Our work paves the way for studying the Bose-Einstein condensations of ultracold gases of microwave-shielded polar molecules.
\end{abstract}

\maketitle

\textit{Introduction}.---Unprecedented advancements have recently been achieved in the field of ultracold molecules~\cite{Ye2019,Luo2021,Luo2022a,Luo2022b,Cao2022,Wang2023,Will2023,Will2023b}. Through dedicated endeavors over a decade~\cite{Ye2008,Nagerl2014,Cornish2014,Zwierlein2015,Bloch2018,Ospelkaus2020,Wang2016, Jamison2017,Will2022,Ni2021,Tarbutt2021,Ye2020a,Ye2021,Ye2020b,Doyle2021}, both degenerate Fermi gases~\cite{Luo2022a} and Bose-Einstein condensates (BECs)~\cite{Will2023b} of polar molecules have been accomplished via microwave shielding techniques. These breakthroughs in microwave-shielded polar molecules (MSPMs) have opened up exceptional avenues for quantum computing and information processing~\cite{DeMille2002,Cornish2020,Zoller2006a,Tesch2002,Wall2015,Albert2020}, quantum
simulation~\cite{Zoller2006,Zwierlein2021}, ultracold chemistry~\cite{Krem2008,Ni2019,Liu2022}, and precision measurement~\cite{Kozlov2007,Berger2010,Hinds2011,Hutzler2020}. Notably, the creation of a degenerate NaK gas~\cite{Luo2022a,Luo2022b} and tetratomic (NaK)$_{2}$ molecules~\cite{chen2023} provides a significant opportunity for realizing topological quantum computing~\cite{Read2000,Ivanov2001} and exploring the intriguing crossover from $p$-wave Bardeen-Cooper-Schrieffer superfluidity~\cite{Pfau2009,You1999,Baranov2002,Shi2010,Pu2010,Hirsch2010, Shlyapnikov2011,Baranov2012,Shi2014,Zhai2013,Deng2023} of molecules to BEC of tetramers.

The interaction potential between two MSPMs is modeled by a long-range dipole-dipole interaction and an $1/r^6$-type short-range shielding core. The typical size of the repulsive core is around $10^3$ Bohr radii~\cite{Luo2022a,Luo2022b,chen2023,Deng2023}, a value comparable to the inter-molecular spacing. These features are in striking contrast to those of the dilute atomic gases for which the range of the two-body potential is much smaller than the inter-atomic spacing. As a result, the atom-atom interaction is completely characterized by the $s$-wave scattering length and condensate is described by the Gross-Pitaevskii equation (GPE). The molecular gases, instead, closely resemble the liquid ${}^4$He for which the size of the repulsive core ($\sim 2\mathrm{\mathring{A}}$) of the Lennard-Jones potential~\cite{Michels1938,Michels1939} is comparable to the inter-atomic distances. Such strong repulsion prevents the establishment of coherence within the inter-particle distances and results in a small condensate fraction $\sim 10\%$~\cite{Onsager,Sears1982,Sears1983,Sokol1990,Sokol1991}. Numerous theoretical treatments, such as variational
approach~\cite{Jastrow1955,cluster1958,Manousakis1985a,Manousakis1985b,Manousakis1991}, Monte Carlo (MC) simulations~\cite{McMillan1965,GFMC,PIMC,DiffMC}, and the backflow approach~\cite{Feynman1956}, have been developed for tackling liquid $^4$He. Remarkably, MSPMs feature both longer-range attractive and repulsive interactions than the Lennard-Jones potential, giving rise to more intriguing many-body correlated phenomena, e.g., molecular droplets stabilized by the two-body shielding potential. The recent achievement in NaCs BEC~\cite{Will2023b} offers a natural platform for delving into these strongly correlated phenomena beyond the realm of atomic BEC and $^{4}$He physics.

In this Letter, we study the ground-state properties of the ultracold gases of bosonic MSPMs by taking into account the two-body correlation. Based on a variational wavefunction with Jastow correlations (VWJC)~\cite{Jastrow1955,Shi2018}, we derive, using cluster expansion~\cite{cluster1958}, an expression for the total energy which can be minimized to yield the many-body wavefunctions. For untrapped gases of MSPMs, we show that the system is always stable and supports self-bound gas (SBG) and expanding gas (EG) phases, which is in striking contrast to the results of the GPE approach. We also compute the condensate fraction, which shows that a large number molecules may be depleted out of the condensate when the size of the shielding core of the two-body potential becomes comparable to the inter-molecular distance. These results distinguish the molecular condensates from the atomic ones and invalidate the application of the GPE treatment to the ultracold gases of MSPMs. Finally, as to the experimental detection, we calculate the momentum distribution which exhibits the characteristic bimodal feature with condensed molecules occupying the small momentum regime. Given the high tunability and rich measurement techniques, ultracold Bose gases of microwave-shielded polar molecules provide an ideal platform for studying stable condensates with correlation ranging from weak to strong regimes.

\textit{Formulation}.---We consider a gas of $N$ interacting bosons at zero temperature. The Hamiltonian of the system takes the form
\begin{align}
H&=H_{0}+H_{\mathrm{int}},\\
H_{0}&=\int d{\mathbf{r}}\left[ \frac{1}{2M}\nabla \hat{\psi}^{\dagger }({
\mathbf{r}})\nabla \hat{\psi}({\mathbf{r}})+V({\mathbf{r}})\hat{\psi}
^{\dagger }({\mathbf{r}})\hat{\psi}({\mathbf{r}})\right] ,  \nonumber\\
H_{\mathrm{int}}&=\frac{1}{2}\int d{\mathbf{r}}d{\mathbf{r}}^{\prime }U(
\mathbf{r}-\mathbf{r}^{\prime })\hat{\psi}^{\dagger }({\mathbf{r}})\hat{\psi}%
^{\dagger }({\mathbf{r}}^{\prime })\hat{\psi}({\mathbf{r}}^{\prime })\hat{%
\psi}({\mathbf{r}}),  \nonumber
\end{align}%
are, respectively, the single- and two-particle Hamiltonian, $M$ is the mass of the molecule, $\hat{\psi}({\mathbf{r}})$ is the field operator, $V({\mathbf{r}})=M[\omega _{\perp }^{2}(x^{2}+y^{2})+\omega _{z}^{2}z^{2}]/2$ describes the confining potential with $\omega _{\perp }$ and $\omega _{z}$ being the transverse and axial trap frequencies, respectively, and $U({\mathbf{r}})$ is the two-body interaction potential. When the range of the interaction is comparable to the inter-particle distance, one has to take into account the two-body correlation in the many-body wavefunction. To this end, we introduce, based on the coherent state, the variational ansatz~\cite{Shi2018,SM}
\begin{equation*}
\left\vert \Psi \right\rangle =\frac{e^{-\alpha ^{2}/2}}{\sqrt{\mathcal{N}}}%
\sum_{N}\frac{\alpha ^{N}}{N!}\left\vert \Psi _{N}\right\rangle
\end{equation*}%
where $\mathcal{N}$ is a normalization constant, $\alpha$ is a parameter of the coherent state, and the $N$-particle
wavefunction is
\begin{equation}
\left\vert \Psi _{N}\right\rangle =\int D[{\mathbf{r}}]\prod_{i<j(=1)}^{N}J({%
\mathbf{r}}_{i},{\mathbf{r}}_{j})\prod_{j=1}^{N}\phi _{0}({\mathbf{r}}_{j})%
\hat{\psi}^{\dagger }({\mathbf{r}}_{j})\left\vert 0\right\rangle .
\end{equation}%
Here, $\phi _{0}({\mathbf{r}})$ is the normalized single-particle wavefunction and $J({\mathbf{r}}_{i},{\mathbf{r}}_{j})$ $[=J({\mathbf{r}}_{j},{\mathbf{r}}_{i})]$ is the Jastrow correlation factor which vanishes as ${\mathbf{r}}_{i}\rightarrow {\mathbf{r}}_{j}$ and approaches unit as $|{\mathbf{r}}_{i}-{\mathbf{r}}_{j}|\rightarrow \infty $. Interestingly, for conventional BECs with contact interactions, i.e., $J({\mathbf{r}}_{i},{\mathbf{r}}_{j})\approx 1$, the variational wave function reduces to a
coherent state $\left\vert \Psi \right\rangle =e^{-\alpha ^{2}/2}e^{\alpha \hat{b}^{\dagger }}\left\vert 0\right\rangle $, where $\hat{b}=\int d{\mathbf{r}}\phi _{0}({\mathbf{r}})\hat{\psi}({\mathbf{r}})$. While for dense
gases, every boson is surrounded by hole excitations. This many-body hole-dressing effect is captured by the Jastraw-like term $J({\mathbf{r}}_{i},{\mathbf{r}}_{j})$ which gives rise to the non-trivial (normalized)
second-order correlation function
\begin{equation*}
g_{2}({\mathbf{r}},{\mathbf{r}}^{\prime })=\frac{\left\langle \Psi
\right\vert \hat{\psi}^{\dagger }({\mathbf{r}})\hat{\psi}^{\dagger }({%
\mathbf{r}}^{\prime })\hat{\psi}({\mathbf{r}}^{\prime })\hat{\psi}({\mathbf{r%
}})\left\vert \Psi \right\rangle }{n({\mathbf{r}})n({\mathbf{r}}^{\prime })}
\end{equation*}%
in the inter-molecular distance, where $n({\mathbf{r}})=\left\langle \Psi\right\vert \hat{\psi}^{\dagger }({\mathbf{r}})\hat{\psi}({\mathbf{r}})\left\vert \Psi \right\rangle $ is the density distribution.

To determine $J({\mathbf{r}},{\mathbf{r}}^{\prime })$ and the density amplitude $\phi ({\mathbf{r}})=\sqrt{n({\mathbf{r}})}$, we minimize the total energy~\cite{SM},
\begin{align}
E& =\left\langle \Psi \right\vert H\left\vert \Psi \right\rangle\nonumber\\
&=\int d{\mathbf{r}}\left[ \frac{1}{2M}\nabla \phi ({\mathbf{r}})\nabla
\phi ({\mathbf{r}})+V({\mathbf{r}})n({\mathbf{r}})\right]   \notag \\
& \quad +\frac{1}{2}\int d{\mathbf{r}}d{\mathbf{r}}^{\prime }U_{\mathrm{re}}(%
\mathbf{r},\mathbf{r}^{\prime })n({\mathbf{r}})n({\mathbf{r}}^{\prime }),
\end{align}%
where
\begin{align}
U_{\mathrm{re}}(\mathbf{r},\mathbf{r}^{\prime })& =\left\{ U(\mathbf{r}-%
\mathbf{r}^{\prime })J^{2}({\mathbf{r}},{\mathbf{r}}^{\prime })+\frac{1}{M}%
[\nabla J({\mathbf{r}},{\mathbf{r}}^{\prime })]^{2}\right.   \notag \\
& \quad \left. +\frac{1}{4M}\nabla f({\mathbf{r}},{\mathbf{r}}^{\prime
})\cdot \nabla \right\} \bar{g}_{2}({\mathbf{r}},{\mathbf{r}}^{\prime }),
\label{Ure}
\end{align}%
is the two-body interaction renormalized by the hole excitations, $f({\mathbf{r}},{\mathbf{r}}^{\prime })=J^{2}({\mathbf{r}},{\mathbf{r}}^{\prime})-1$, and $\bar{g}_{2}({\mathbf{r}},{\mathbf{r}}^{\prime })=g_{2}({\mathbf{r}},{\mathbf{r}}^{\prime })/J^{2}({\mathbf{r}},{\mathbf{r}}^{\prime })$. In principle, $g_{2}({\mathbf{r}},{\mathbf{r}}^{\prime })$ and $E$ can be evaluated using MC simulations~\cite{McMillan1965}. For recent experiments on bosonic MSPMs~\cite{Wang2023,Will2023b} in low and intermediate density regimes, the inter-molecular spacing is larger than the size of the repulsive core. As a result, we adopt the cluster expansion to calculate
\begin{equation}
\bar{g}_{2}({\mathbf{r}},{\mathbf{r}}^{\prime })=1+\int d{\mathbf{r}}_{1}f(%
\mathbf{r},{\mathbf{r}}_{1})n({\mathbf{r}}_{1})F({\mathbf{r}}_{1},{\mathbf{r}%
}^{\prime })
\end{equation}%
analytically~\cite{cluster1958,SM}, where $F({\mathbf{r}},{\mathbf{r}}^{\prime })=f({\mathbf{r}},{\mathbf{r}}^{\prime })+\int d{\mathbf{r}}_{1}f({\mathbf{r}},{\mathbf{r}}_{1})n({\mathbf{r}}_{1})F({\mathbf{r}}_{1},{\mathbf{r}}^{\prime })$ satisfies the Dyson-like equation.

Since the explicit form of the energy functional $E[\phi ({\mathbf{r}}),J({\mathbf{r}},{\mathbf{r}}^{\prime })]$ leads to $\delta E/\delta \phi ({\mathbf{r}})$ and $\delta E/\delta J({\mathbf{r}},{\mathbf{r}}^{\prime })$
analytically, it is convenient to minimize the energy and determine the ground state configuration $\{\phi ({\mathbf{r}}),J({\mathbf{r}},{\mathbf{r}}^{\prime })\}$ via the gradient descent algorithm. Compared to the typical MC calculations, the variational treatments are computationally inexpensive. And more importantly, because the molecular gases are very dilute, the cluster expansion calculation is more reliable here than in the liquid $^4$He.

\begin{figure}[tbp]
\centering
\includegraphics[trim=0 0 90 5, clip, width=1\columnwidth]{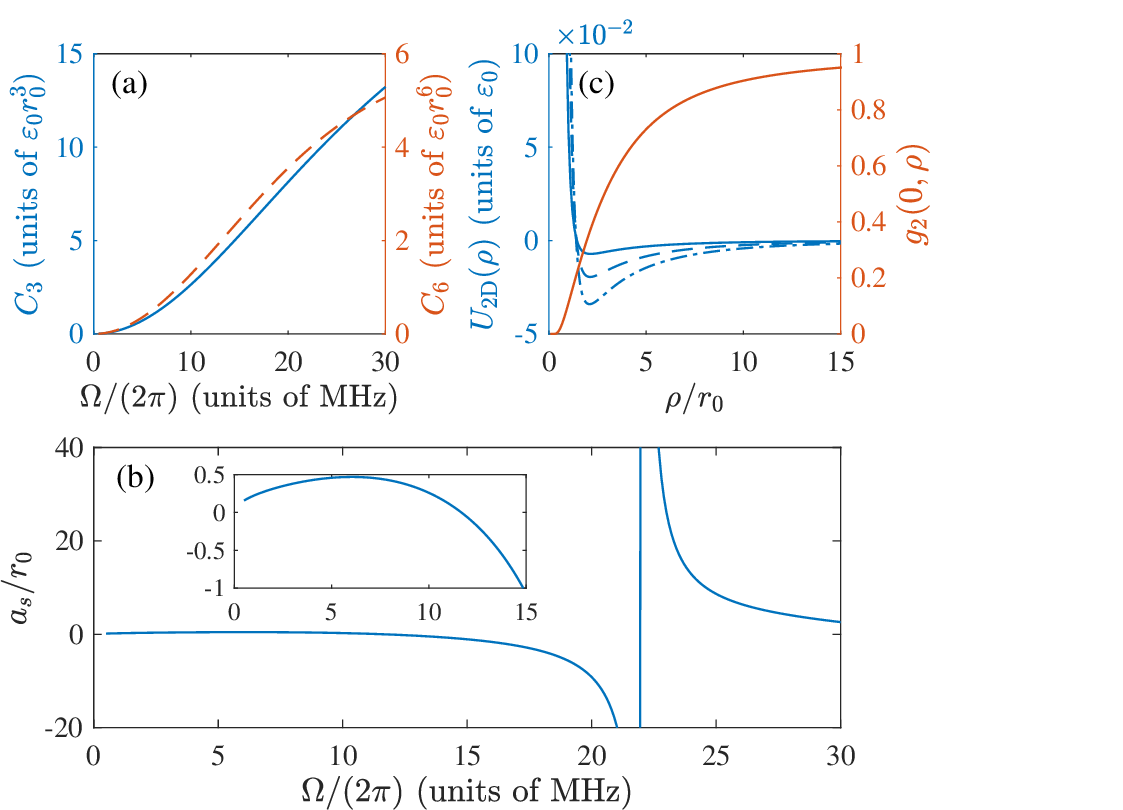}
\caption{(color online). Two-body interactions for MWS NaRb molecules with microwave detuning $\Delta/(2\protect\pi)=30\,\mathrm{MHz}$. (a) $C_3$ and $C_6$ versus $\Omega$. (b) $s$-wave scattering length $a_s$ versus $\Omega$. The inset is the zoom-in plot for $a_s$ for small $\Omega$. (c) Left $y$ axis (blue) shows $U_{\mathrm{2D}}(\protect\rho)$ for $\Omega/(2\protect\pi)=8$, $14$, $20\,\mathrm{MHz}$ (in descending order of minimal values of the potentials). Right $y$ axis (red) shows the typical $g_2(0,\rho)$.}
\label{potential}
\end{figure}

\textit{Ultracold gases of MSPMs}.---For MSPMs, the effective inter-molecular potential takes the form~\cite{Deng2023}
\begin{equation}
U(\mathbf{r})=\frac{C_{3}}{r^{3}}(3\cos ^{2}\theta -1)+\frac{C_{6}}{r^{6}}%
\sin ^{2}\theta (1+\cos ^{2}\theta ),
\end{equation}%
where $\theta$ is the polar angle of $\mathbf{r}$, and $C_{3}$ and $C_{6}$ are interaction strengths tunable via the Rabi frequency $\Omega$ and detuning $\Delta$ of the microwave field. Apparently, the $C_3$ term is a negated dipolar interaction that is attractive on the $xy$ plane and repulsive along the $z$ axis. The $C_6$ term represents the short-range shielding potential. As a concrete example, we present, in Fig. \ref{potential}(a), $C_{3}$ and $C_{6}$ of NaRb molecule as a function of $\Omega$. Here and henceforth, we fix the detuning of the microwave at $\Delta/(2\pi)=30\,\mathrm{MHz}$. For convenience, we introduce $r_0\equiv10^3a_0$ (with $a_0$ being the Bohr radius) and $\varepsilon_0\equiv\hbar^2/(Mr_0^2)$ as the units for length and energy, respectively. The typical size of the shielding core is around $r_0$. We remark that even higher tunability of the interactions can be achieved by introducing a $\pi$-polarized microwave as in the recent experiment~\cite{Will2023b}.

To compare the VWJC results with those obtained by GPE, we numerically compute the $s$-wave scattering length, $a_s$, of the potential $U({\mathbf{r}})$. Then the pseudo-potential in the GPE approach is modeled as~\cite{Yi2000,Yi2001}
\begin{align}
U_{\mathrm{pp}}({\mathbf{r}})=\frac{4\pi\hbar^2 a_s}{M}\delta({\mathbf{r}})+%
\frac{C_3}{r^3}(3\cos^2\theta-1).
\end{align}
Figure~\ref{potential}(b) shows $a_{s}$ as a function of $\Omega$, on which a scattering resonance emerges at $\Omega/(2\pi)\approx 22\,\mathrm{MHz}$. It is now clear that both $U({\mathbf{r}})$ and $U_{\mathrm{pp}}({\mathbf{r}}
)$ are completely determined once the physical parameters $\Omega$ and $\Delta$ are given.

We concentrate on the quasi-2D geometries, where the axial trap frequency is sufficiently strong such that the motion of the molecules along the $z$ axis is frozen to the ground state of the axial harmonic oscillator, i.e., $\phi_z(z)=e^{-z^{2}/(2a_{z}^{2})}/(\pi ^{1/4}a_{z}^{1/2})$ with $a_z=\sqrt{\hbar/(M\omega_z)}$. As a result, the field operator can be factorized into $\hat{\psi}({\mathbf{r}})=\hat\psi({\boldsymbol{\rho}})\phi_z(z)$, where ${\boldsymbol{\rho}}=(x,y)$. After integrating out the $z$ variable from the Hamiltonian, the system reduced to a quasi-2D one through the following replacements: ${\mathbf{r}}\rightarrow{\boldsymbol{\rho}}$, $V\rightarrow V_{\mathrm{2D}}({\boldsymbol{\rho}})=M\omega_\perp^2\rho^2/2$, and $U\rightarrow U_{2D}({\boldsymbol{\rho}}-{\boldsymbol{\rho}}^{\prime})=\int dzdz^{\prime}U(\mathbf{r}-\mathbf{r}^{\prime}) \phi_{z}^{2}(z)\phi_{z}^{2}(z^{\prime})$. In Fig. \ref{potential}(c), we plot $U_{2D}(\rho)$ for different $\Omega$'s~\cite{SM}. Generally, $U_{\mathrm{2D}}$ is attractive (repulsive) in the long (short) range. And the
potential well of $U_{\mathrm{2D}}$ becomes deeper as $\Omega$ increases.

Below, we explore the ground-state properties of the ultracold molecular gases. The axial trap frequency is fixed at  $\omega_z/(2\pi)=1.3\,{\rm kHz}$. Moreover, we also solve GPE with the pseudo-potential $U_{\mathrm{pp}}({\mathbf{r}})$ to illustrate drastic distinctions between molecular gases and atomic BECs.

\begin{figure}[tbp]
\centering
\includegraphics[width=1\columnwidth]{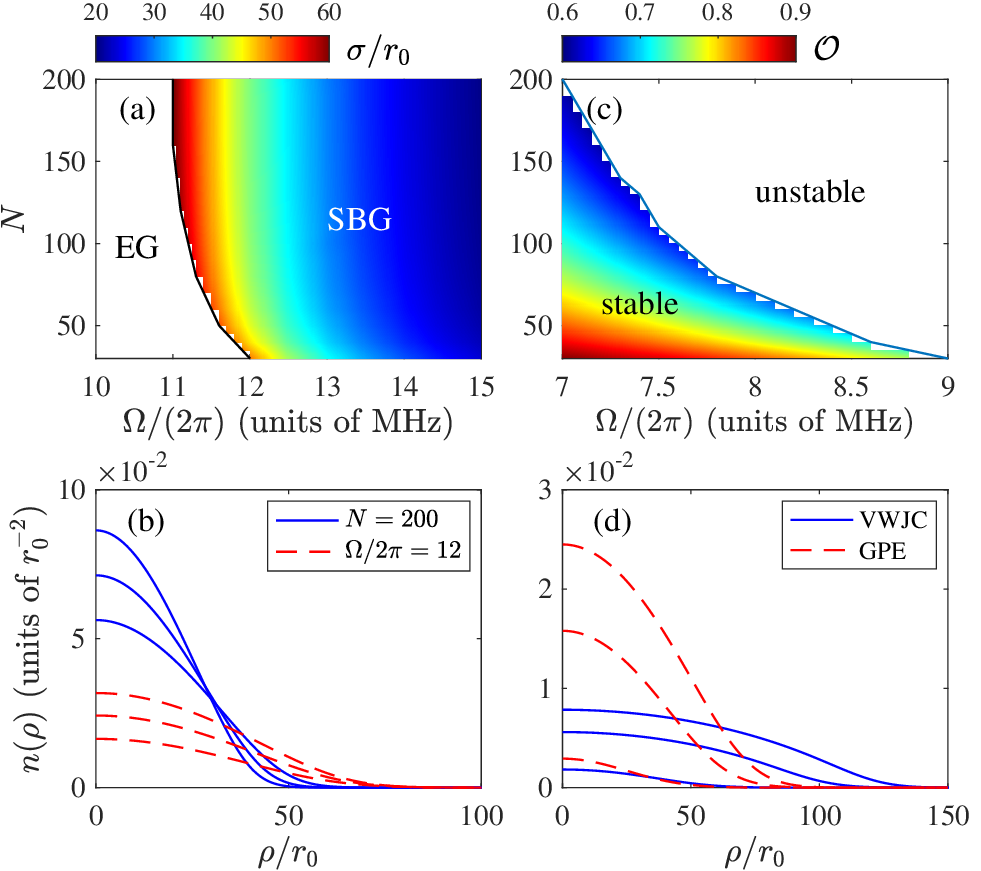}
\caption{(color online). (a) Phase diagram obtained via VWJC. The solid line marks the boundary between the EG and SBG phases. The colormap in SBG represents the radial size of the condensate. (b) Total densities in the SBG phase. In descending order of the central densities, the solid lines are $N=200$ with $\Omega/(2\protect\pi)=14$, $13.5$, and $13\,\mathrm{MHz}$ and the dashed lines are $\Omega/(2\protect\pi)=12\,\mathrm{MHz}$ with $N=200$, $150$, and $100$. (c) Stability diagram of the trapped molecular BECs obtained via GPE. The colormap in the stable regime is the overlap between the densities obtained via GPE and VWJC. (d) Comparison of the total densities obtained via VWJC (solid lines) and GPE (dashed lines) for $\Omega/(2\protect\pi)=5\,\mathrm{MHz}$ and $N=200$, $100$, and $10$ (in descending order of the central densities).}
\label{phase}
\end{figure}

\textit{Quantum phases}.---In Fig.~\ref{phase}(a), we map out the phase diagram for the untrapped molecular gases on the $\Omega$-$N$ parameter plane. When the attractive interaction is sufficient strong, the molecules form SBG which has a finite radial size, $\sigma=\left[2\pi N^{-1}\int d\rho \rho^3n(\rho)\right]^{1/2}$, even in the absence of the external traps. Otherwise, the molecules become an EG which expands without external traps. The color map in the SBG phases of Fig.~\ref{phase}(a) represents the radial size of the BECs which indicates that the gases become more tightly bound by increasing either $N$ or $\Omega$. To reveal more details of the SBG phase, we plot, in Fig.~\ref{phase}(b), the total density of the gases in the SBG phase. For a fixed $N$, the radial size of the condensates rapidly decreases as $\Omega$ is increased; while when $\Omega$ is fixed, increasing $N$ does not significantly modify the profile of the density.

For the GPE treatment, it turns out that, in the absence of external trap, the molecular gases either expand or collapse. To fix this, we impose a transverse trap with frequency $\omega_\perp/(2\pi)=36\,\mathrm{Hz}$ in all GPE calculations (and also in the corresponding VWJC calculations). In Fig.~\ref{phase}(c), we present the stability diagram of the gases obtained by solving GPE. Although the VWJC solutions of the trapped gases are always stable, the GPE approach shows that the condensate may collapse even when the interaction is relatively weak. The color map in the stable regime represents the overlap between the GPE density ($n_{\mathrm{GPE}}$) and VWJC density, i.e., $\mathcal{O}=2\pi N^{-1}\int \rho d\rho [n(\rho)n_{\mathrm{GPE}}(\rho)]^{1/2}$. We also compare these two densities directly in Fig.~\ref{phase}(d). As can be seen, $n(\rho)$ and $n_{\mathrm{GPE}}(\rho)$ are comparable only when the interactions are very weak (e.g., $\Omega=2\pi\times5\,\mathrm{MHz}$ and $N=10$). While for typical molecular densities, the discrepancy of two approaches can be quite large. In fact, since the size of the large repulsive core is ignored in the GPE treatment, the width of $n_{\mathrm{GPE}}(\rho)$ is significantly smaller than that of $n(\rho)$. These results manifest that the GPE treatment is generally inapplicable for the ultracold gases of MSPMs.

To gain more insight into the ground-state properties, we plot the typical second-order correlation function $g_{2}(0,\rho)$ in Fig.~\ref{potential}(c). Remarkably, $g_2$ saturates to unit at a distance comparable to the size of the whole molecular gases, suggesting the existence of a strong many-body correlation among all molecules. This is in striking contrast to the situation in liquid $^4$He and is likely due to the long-range shielding potential of MSPMs.

\begin{figure}[tbp]
\centering
\includegraphics[width=1\columnwidth]{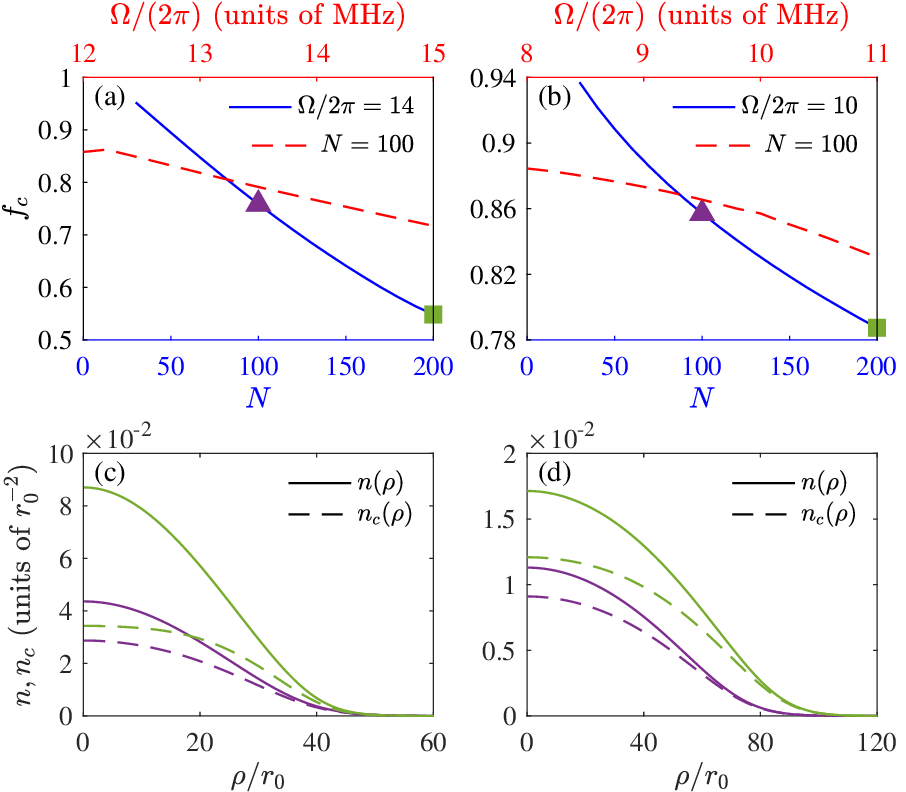}
\caption{(color online). Condensation fraction versus $N$ and $\Omega$ in the SBG phase (a) and EG phase (b). (c) Total (solid line) and condensate (dashed line) densities for parameters denoted by the markers in (a). (d)
Total (solid line) and condensate (dashed line) densities for parameters denoted by the markers in (b).}
\label{condfrac}
\end{figure}

\textit{Condensation fraction.}---Since the strong repulsion can deplete molecules out of the condensate, it is necessary to examine the condensate fraction in molecular gases. To this end, we compute the first-order correlation function $G_{1}({\boldsymbol{\rho}},{\boldsymbol{\rho}}^{\prime})=\left\langle \Psi \right\vert \hat{\psi}^{\dagger}({\boldsymbol{\rho}}^{\prime })\hat{\psi}({\boldsymbol{\rho}})\left\vert \Psi \right\rangle$~\cite{SM}. After numerically diagonalizing $G_1$, we obtain $G_{1}({\boldsymbol{\rho}},{\boldsymbol{\rho}}^{\prime })=\sum_{\ell\geq0}N_\ell\bar\varphi_\ell({\boldsymbol{\rho}}) \bar\varphi_\ell^*({\boldsymbol{\rho}}^{\prime })$, where the occupation number $N_\ell$ in the normalized mode $\bar\varphi_\ell({\boldsymbol{\rho}})$ is sorted in descending order. Then $N_0$ is the number of molecules in the condensation and $f_c=N_0/N$ is the condensate fraction.

In Fig.~\ref{condfrac}(a) and (b), we plot the $\Omega $ and $N$ dependence of $f_{c}$ for the molecular gases in the SBG and EG phases, respectively. In general, the condensate fraction in the SBG phase is smaller than that in the EG phase due to the higher density and thus stronger repulsion. And in both cases, $f_{c}$ decreases with the increase of either $\Omega $ or $N$. Particularly, in the SBG regime ($\Omega =2\pi \times 14\,\mathrm{MHz}$), the condensation fraction can be as low as $0.55$ for $N=200$. Although the molecular gases are dilute in the sense of absolute density, they remain to be strongly interacting since the inter-molecular distance is comparable to the size of the shielding core. These results further invalidate the GPE treatments for condensates of MSPMs. Figure~\ref{condfrac}(c) and (d) compare the total density $n(\rho )$ and the condensate density $n_{c}(\rho) =N_{0}|\bar{\varphi}_{0}(\rho )|^{2}$ in the SBG and EG regimes, respectively. In analogy to $n(\rho )$, the depletion density $(n-n_{c})$ is also a decreasing function of the radius $\rho$, which indicates that higher total density leads to higher depletion density.

\begin{figure}[tbp]
\centering
\includegraphics[width=1\columnwidth]{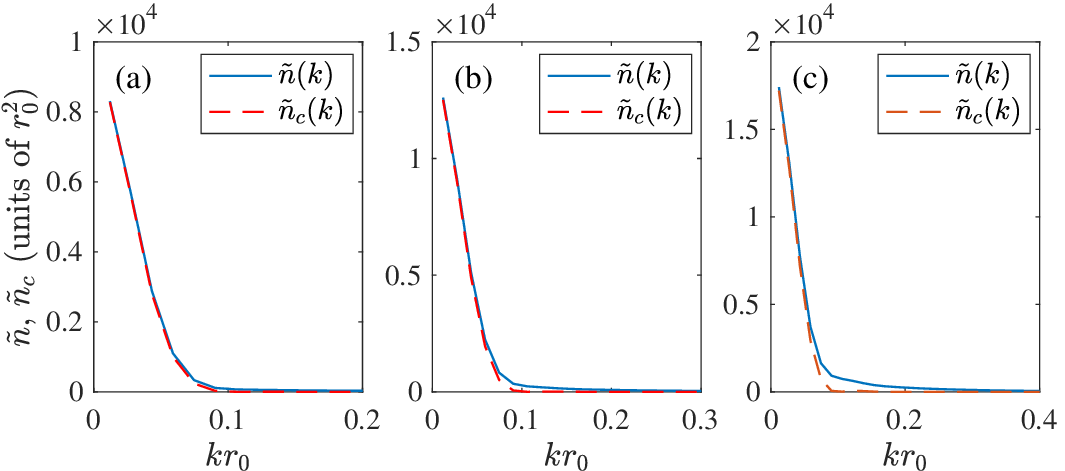}
\caption{$\tilde n(k)$ and $\tilde n_c(k)$ for $\Omega/(2\protect\pi)=14\,\mathrm{MHz}$ with $N=50$ (a), $100$ (b), and $200$ (c).}
\label{moment}
\end{figure}

\textit{Momentum distributions}.---As to the experimental detection of the molecular condensates, we explore the momentum distribution of the gases. For this purpose, we note that the momentum distribution of the whole molecular gas is $\tilde n(\mathbf{k})=(2\pi)^{-2}\int d{\boldsymbol{\rho}}d{\boldsymbol{\rho}}^{\prime }e^{i\mathbf{k}\cdot {\boldsymbol{\rho}}^{\prime}-i\mathbf{k}\cdot {\boldsymbol{\rho}}}G_{1}({\boldsymbol{\rho}}, {\boldsymbol{\rho}}^{\prime })$; while for condensed molecules, we have $\tilde n_c(\mathbf{k})=(2\pi)^{-2}N_0|\int d^{2}{\boldsymbol{\rho}}e^{-i\mathbf{k}\cdot {\boldsymbol{\rho}}}\bar\varphi_0({\boldsymbol{\rho}})|^2$. Figure~\ref{moment}(a)-(c) plot the momentum distributions of all and condensed molecules for a fixed $\Omega/(2\pi)=14\,\mathrm{MHz}$ with $N=50$, $100$, and $200$, respectively. The corresponding condensate fractions
are, respectively, $f_c=0.89$, $0.74$, and $0.55$. It is seen that the condensed molecules are dominant for small $k$ and the uncondensed molecules occupy the large $k$ region. Consequently, $\tilde n(k)$ generally exhibits a bimodal density distribution even at zero temperature. Finally, taking into account the thermal molecules, the experimentally measured momentum distributions may be of a trimodal type, which can be used as signature for the Bose condensation of molecules.

\textit{Conclusion}.---We have studied the ground-state properties of the ultracold gases of MSPMs by means of both VWJC and GPE approaches. In contrast to the GPE results, the VWJC calculations show that the system is always stable and supports SBG and EG phase in the absence of external trap. More importantly, we have shown that the simple GPE is inapplicable to molecular condensates of MSPMs with typical gas densities. Instead, the two-body correlation must be considered either variationally through Jastrow factor or, more reliably, using quantum MC methods. Although our studies focus on NaRb molecule, the approach can be straightforwardly applied to the calculation of NaCs molecule. In fact, due to reduced dipolar interaction by the $\pi$-polarized microwave, our calculations show that the typical condensate fractions for EG and SBG phases in NaCs experiments~\cite{Will2023b} are smaller than $78\%$.

This work was supported by the NSFC (Grants No. 12135018, No.12047503, and No. 12274331) and by National Key Research and Development Program of China (Grant No. 2021YFA0718304).

\bibliography{ref_dMolecule.bib}

\clearpage

\widetext

\begin{center}
\textbf{\large Supplemental Materials: }
\end{center}

This Supplemental Material is structured as follows. In the first section,
we introduce the variational ansatz and provide a generic formula for the ground state energy in
two different manners. Using the chain diagram resummation in the cluster expansion, we obtain the analytic expression of the energy. In the second section, we derive the effective interaction in quasi-2D. By diagonalizing the first order correlation function, we determine the condensate wavefunction and the condensate fraction. The momentum distribution is also obtained using the Fourier transform of the first order correlation function, which can be measured in time-of-flight experiments.

\setcounter{equation}{0} \setcounter{figure}{0} \setcounter{table}{0} %
\setcounter{page}{1} \setcounter{section}{0} \makeatletter%
\renewcommand{\theequation}{S\arabic{equation}} \renewcommand{\thefigure}{S%
\arabic{figure}} \renewcommand{\bibnumfmt}[1]{[S#1]} \renewcommand{%
\citenumfont}[1]{S#1} \renewcommand{\thesection}{S\arabic{section}}%
\setcounter{secnumdepth}{3}

\section{Non-Gaussian variational ansatz \label{sm:NGS}}

In this section, we introduce the non-Gaussian variational ansatz, and
calculate the ground state energy. The variational ansatz can be
re-expressed in a compact form $\left\vert \Psi \right\rangle =S\left\vert
\alpha \right\rangle /\sqrt{\mathcal{N}}$, i.e., the non-Gaussian similar
transformation \cite{Shi2018}%
\begin{equation}
S=\exp [\frac{1}{2}\int d\mathbf{r}d\mathbf{r}^{\prime }\chi (\mathbf{r,r}%
^{\prime })\hat{\psi}^{\dagger }(\mathbf{r})\hat{\psi}^{\dagger }(\mathbf{r}%
^{\prime })\hat{\psi}(\mathbf{r}^{\prime })\hat{\psi}(\mathbf{r})]
\end{equation}%
acting on the coherent state $\left\vert \alpha \right\rangle =e^{\alpha
(b^{\dagger }-b)}\left\vert 0\right\rangle $, where $b=\int d\mathbf{r}\phi
_{0}(\mathbf{r})\hat{\psi}(\mathbf{r})$ and $J(\mathbf{r}_{i},\mathbf{r}%
_{j})=e^{\chi (\mathbf{r}_{i},\mathbf{r}_{j})}$. The ansatz $\left\vert \Psi
\right\rangle $ does not conserve the particle number, thus, it breaks the $%
U(1)$ symmetry. Alternatively, by projecting on a certain particle number
sector, one can also employ the ansatz $\left\vert \Psi _{N}\right\rangle =%
\mathcal{P}_{N}\left\vert \Psi \right\rangle $:%
\begin{equation}
\left\vert \Psi _{N}\right\rangle =\frac{1}{\sqrt{\mathcal{N}}}\int D[%
\mathbf{r}]\prod_{i<j(=1)}^{N}J(\mathbf{r}_{i},\mathbf{r}_{j})%
\prod_{j=1}^{N}\phi _{0}(\mathbf{r}_{j})\hat{\psi}^{\dagger }(\mathbf{r}%
_{j})\left\vert 0\right\rangle
\end{equation}%
to describe the ground state properties. In the following, we shall derive
the same expression of the ground state energy for the two variational
states. Hereafter, we consider the ground state without the time-reversal
symmetry breaking, namely, $\phi _{0}$ and $J(\mathbf{r}_{i},\mathbf{r}_{j})$
are real.

\subsection{Variational energy for the state $\left\vert \Psi \right\rangle $%
}

For the first ansatz, the normalization factor, the first order and second
order correlation functions are%
\begin{equation}
\mathcal{N}=e^{-\alpha ^{2}}\sum_{N}\frac{\alpha ^{2N}}{N!}\int D[\mathbf{r}%
]\prod_{i<j=1}^{N}J^{2}(\mathbf{r}_{i},\mathbf{r}_{j})\prod_{j=1}^{N}\phi
_{0}^{2}(\mathbf{r}_{j}),
\end{equation}%
\begin{eqnarray}
G_{1}(\mathbf{r},\mathbf{r}^{\prime }) &=&\left\langle \Psi \right\vert \hat{%
\psi}^{\dagger }(\mathbf{r}^{\prime })\hat{\psi}(\mathbf{r})\left\vert \Psi
\right\rangle   \notag \\
&=&\alpha ^{2}\phi _{0}(\mathbf{r})\phi _{0}(\mathbf{r}^{\prime })\frac{%
e^{-\alpha ^{2}}}{\mathcal{N}}\sum_{N}\frac{\alpha ^{2N}}{N!}\int D[\mathbf{r%
}]\prod_{i<j=1}^{N}J^{2}(\mathbf{r}_{i},\mathbf{r}_{j})\prod_{j=1}^{N}[\phi
_{0}^{2}(\mathbf{r}_{j})J(\mathbf{r},\mathbf{r}_{j})J(\mathbf{r}^{\prime },%
\mathbf{r}_{j})],
\end{eqnarray}%
and%
\begin{eqnarray}
G_{2}(\mathbf{r},\mathbf{r}^{\prime }) &=&\left\langle \Psi \right\vert \hat{%
\psi}^{\dagger }(\mathbf{r})\hat{\psi}^{\dagger }(\mathbf{r}^{\prime })\hat{%
\psi}(\mathbf{r}^{\prime })\hat{\psi}(\mathbf{r})\left\vert \Psi
\right\rangle   \notag \\
&=&\alpha ^{4}\phi _{0}^{2}(\mathbf{r})\phi _{0}^{2}(\mathbf{r}^{\prime
})J^{2}(\mathbf{r},\mathbf{r}^{\prime })\frac{e^{-\alpha ^{2}}}{\mathcal{N}}%
\sum_{N}\frac{\alpha ^{2N}}{N!}\int D[\mathbf{r}]\prod_{i<j=1}^{N}J^{2}(%
\mathbf{r}_{i},\mathbf{r}_{j})\prod_{j=1}^{N}[\phi _{0}^{2}(\mathbf{r}%
_{j})J^{2}(\mathbf{r},\mathbf{r}_{j})J^{2}(\mathbf{r}^{\prime },\mathbf{r}%
_{j})].
\end{eqnarray}

We notice that in $\mathcal{N}$ and $G_{1,2}(\mathbf{r},\mathbf{r}^{\prime })
$ the integrals have the same pattern, i.e., $\int D[\mathbf{r}%
]\prod_{i<j=1}^{N}J^{2}(\mathbf{r}_{i},\mathbf{r}_{j})\prod_{j=1}^{N}O(%
\mathbf{r}_{j})$. By defining $f(\mathbf{r}_{i},\mathbf{r}_{j})=J^{2}(%
\mathbf{r}_{i},\mathbf{r}_{j})-1$, we can use the standard cluster expansion
\cite{cluster1958} to obtain%
\begin{eqnarray}
\mathcal{N} &=&\exp [\frac{1}{2!}\alpha ^{4}\int d\mathbf{r}_{1}d\mathbf{r}%
_{2}f(\mathbf{r}_{1},\mathbf{r}_{2})\phi _{0}^{2}(\mathbf{r}_{1})\phi
_{0}^{2}(\mathbf{r}_{2})+...],  \notag \\
G_{1}(\mathbf{r},\mathbf{r}^{\prime }) &=&\alpha ^{2}\phi _{0}(\mathbf{r}%
)\phi _{0}(\mathbf{r}^{\prime })\frac{e^{-\alpha ^{2}}}{\mathcal{N}}\exp
[\alpha ^{2}\int d\mathbf{r}_{1}J(\mathbf{r},\mathbf{r}_{1})J(\mathbf{r}%
^{\prime },\mathbf{r}_{1})\phi _{0}^{2}(\mathbf{r}_{1})  \notag \\
&&+\frac{1}{2!}\alpha ^{4}\int d\mathbf{r}_{1}d\mathbf{r}_{2}f(\mathbf{r}%
_{1},\mathbf{r}_{2})\prod_{j=1,2}J(\mathbf{r},\mathbf{r}_{j})J(\mathbf{r}%
^{\prime },\mathbf{r}_{j})\phi _{0}^{2}(\mathbf{r}_{j})+...],  \notag \\
G_{2}(\mathbf{r},\mathbf{r}^{\prime }) &=&\alpha ^{4}\phi _{0}^{2}(\mathbf{r}%
)\phi _{0}^{2}(\mathbf{r}^{\prime })J^{2}(\mathbf{r},\mathbf{r}^{\prime })%
\frac{e^{-\alpha ^{2}}}{\mathcal{N}}\exp [\alpha ^{2}\int d\mathbf{r}%
_{1}\phi _{0}^{2}(\mathbf{r}_{1})J^{2}(\mathbf{r},\mathbf{r}_{1})J^{2}(%
\mathbf{r}^{\prime },\mathbf{r}_{1})  \notag \\
&&+\frac{1}{2!}\alpha ^{4}\int d\mathbf{r}_{1}d\mathbf{r}_{2}f(\mathbf{r}%
_{1},\mathbf{r}_{2})\prod_{j=1,2}J^{2}(\mathbf{r},\mathbf{r}_{j})J^{2}(%
\mathbf{r}^{\prime },\mathbf{r}_{j})\phi _{0}^{2}(\mathbf{r}_{j})+...].
\end{eqnarray}

The density $n(\mathbf{r})=G_{1}(\mathbf{r},\mathbf{r})$ can be related to $%
\phi _{0}^{2}(\mathbf{r})$ and $f(\mathbf{r},\mathbf{r}^{\prime })$ as%
\begin{eqnarray}
n(\mathbf{r}) &=&\alpha ^{2}\phi _{0}^{2}(\mathbf{r})\exp [\alpha ^{2}\int d%
\mathbf{r}_{1}f(\mathbf{r},\mathbf{r}_{1})\phi _{0}^{2}(\mathbf{r}%
_{1})+\alpha ^{4}\int d\mathbf{r}_{1}d\mathbf{r}_{2}f(\mathbf{r},\mathbf{r}%
_{1})f(\mathbf{r}_{1},\mathbf{r}_{2})\prod_{j=1,2}\phi _{0}^{2}(\mathbf{r}%
_{j})  \notag \\
&&+\frac{1}{2!}\alpha ^{4}\int d\mathbf{r}_{1}d\mathbf{r}_{2}f(\mathbf{r},%
\mathbf{r}_{1})f(\mathbf{r}_{1},\mathbf{r}_{2})f(\mathbf{r}_{2},\mathbf{r}%
)\prod_{j=1,2}\phi _{0}^{2}(\mathbf{r}_{j})+...].  \label{den}
\end{eqnarray}%
The physical relevant quantity is the density $n(\mathbf{r})$ rather than
the bare wavefunction $\phi _{0}(\mathbf{r})$ in the ansatz, therefore, we
invert Eq. (\ref{den}) and express $\alpha ^{2}\phi _{0}^{2}(\mathbf{r})$ in
terms of $n(\mathbf{r})$ and $f(\mathbf{r},\mathbf{r}^{\prime })$ as%
\begin{eqnarray}
\alpha ^{2}\phi _{0}^{2}(\mathbf{r}) &=&n(\mathbf{r})\exp [-\int d\mathbf{r}%
_{1}f(\mathbf{r},\mathbf{r}_{1})n(\mathbf{r}_{1})  \notag \\
&&-\frac{1}{2!}\int d\mathbf{r}_{1}d\mathbf{r}_{2}f(\mathbf{r},\mathbf{r}%
_{1})f(\mathbf{r}_{1},\mathbf{r}_{2})f(\mathbf{r}_{2},\mathbf{r}%
)\prod_{j=1,2}n(\mathbf{r}_{j})+...].  \label{n0vsn}
\end{eqnarray}%
Substituting $\alpha ^{2}\phi _{0}^{2}(\mathbf{r})$ via Eq. (\ref{n0vsn}),
we obtain%
\begin{eqnarray}
G_{1}(\mathbf{r},\mathbf{r}^{\prime }) &=&\phi (\mathbf{r})\phi (\mathbf{r}%
^{\prime })\exp \{\int d\mathbf{r}_{1}[J(\mathbf{r},\mathbf{r}_{1})J(\mathbf{%
r}^{\prime },\mathbf{r}_{1})-1]n(\mathbf{r}_{1})-\frac{1}{2}\int d\mathbf{r}%
_{1}[f(\mathbf{r},\mathbf{r}_{1})+f(\mathbf{r}^{\prime },\mathbf{r}_{1})]n(%
\mathbf{r}_{1})  \notag \\
&&+\frac{1}{2!}\int d\mathbf{r}_{1}d\mathbf{r}_{2}f(\mathbf{r}_{1},\mathbf{r}%
_{2})\prod_{j=1,2}[J(\mathbf{r},\mathbf{r}_{j})J(\mathbf{r}^{\prime },%
\mathbf{r}_{j})-1]\prod_{j=1,2}n(\mathbf{r}_{j})  \notag \\
&&-\frac{1}{4}\int d\mathbf{r}_{1}d\mathbf{r}_{2}f(\mathbf{r}_{1},\mathbf{r}%
_{2})[f(\mathbf{r},\mathbf{r}_{1})f(\mathbf{r}_{2},\mathbf{r})+f(\mathbf{r}%
^{\prime },\mathbf{r}_{1})f(\mathbf{r}_{2},\mathbf{r}^{\prime
})]\prod_{j=1,2}n(\mathbf{r}_{j})+...\},  \label{G1}
\end{eqnarray}%
and%
\begin{eqnarray}
G_{2}(\mathbf{r},\mathbf{r}^{\prime }) &=&n(\mathbf{r})n(\mathbf{r}^{\prime
})J^{2}(\mathbf{r},\mathbf{r}^{\prime })\exp \{\int d\mathbf{r}_{1}f(\mathbf{%
r},\mathbf{r}_{1})n(\mathbf{r}_{1})f(\mathbf{r}_{1},\mathbf{r}^{\prime })
\notag \\
&&+\int d\mathbf{r}_{1}d\mathbf{r}_{2}f(\mathbf{r}_{1},\mathbf{r}%
_{2})\prod_{j=1,2}n(\mathbf{r}_{j})[f(\mathbf{r},\mathbf{r}_{1})f(\mathbf{r}%
_{2},\mathbf{r}^{\prime })+f(\mathbf{r},\mathbf{r}_{1})f(\mathbf{r},\mathbf{r%
}_{2})f(\mathbf{r}^{\prime },\mathbf{r}_{1})  \notag \\
&&+f(\mathbf{r},\mathbf{r}_{1})f(\mathbf{r}^{\prime },\mathbf{r}_{1})f(%
\mathbf{r}^{\prime },\mathbf{r}_{2})+\frac{1}{2}f(\mathbf{r},\mathbf{r}%
_{1})f(\mathbf{r},\mathbf{r}_{2})f(\mathbf{r}^{\prime },\mathbf{r}_{1})f(%
\mathbf{r}^{\prime },\mathbf{r}_{2})]+...\},  \label{G2}
\end{eqnarray}%
as functions of $\phi (\mathbf{r})$ and $J(\mathbf{r},\mathbf{r}^{\prime })$%
, where $n(\mathbf{r})=\phi ^{2}(\mathbf{r})$.

The free-particle energy%
\begin{equation}
E_{0}=\left\langle \Psi \right\vert H_{0}\left\vert \Psi \right\rangle =\int
d\mathbf{r}[\frac{1}{2M}\lim_{\mathbf{r}^{\prime }\rightarrow \mathbf{r}%
}\nabla ^{\prime }\nabla G_{1}(\mathbf{r},\mathbf{r}^{\prime })+V(\mathbf{r}%
)n(\mathbf{r})]
\end{equation}%
and the interaction energy%
\begin{equation}
E_{\mathrm{int}}=\left\langle \Psi \right\vert H_{\mathrm{int}}\left\vert
\Psi \right\rangle =\frac{1}{2}\int d\mathbf{r}d\mathbf{r}^{\prime }U(%
\mathbf{r-r}^{\prime })G_{2}(\mathbf{r},\mathbf{r}^{\prime })
\end{equation}%
can be related to $G_{1}(\mathbf{r},\mathbf{r}^{\prime })$ and $G_{2}(%
\mathbf{r},\mathbf{r}^{\prime })$. Using Eqs. (\ref{G1}) and (\ref{G2}), we
obtain $E=\left\langle \Psi \right\vert H\left\vert \Psi \right\rangle $:%
\begin{eqnarray}
E &=&\int d\mathbf{r}[\frac{1}{2M}\nabla \phi (\mathbf{r})\nabla \phi (%
\mathbf{r})+V(\mathbf{r})n(\mathbf{r})]+\frac{1}{2}\int d\mathbf{r}d\mathbf{r%
}^{\prime }U(\mathbf{r-r}^{\prime })G_{2}(\mathbf{r},\mathbf{r}^{\prime })
\notag \\
&&+\frac{1}{2M}\int d\mathbf{r}d\mathbf{r}^{\prime }n(\mathbf{r})n(\mathbf{r}%
^{\prime })[\nabla J(\mathbf{r},\mathbf{r}^{\prime })\nabla J(\mathbf{r},%
\mathbf{r}^{\prime })+\frac{1}{4}\nabla f(\mathbf{r},\mathbf{r}^{\prime
})\nabla ]\bar{g}_{2}(\mathbf{r},\mathbf{r}^{\prime }),  \label{E1}
\end{eqnarray}%
or equivalently%
\begin{eqnarray}
E &=&\int d\mathbf{r}[\frac{1}{2M}\nabla \phi (\mathbf{r})\nabla \phi (%
\mathbf{r})+V(\mathbf{r})n(\mathbf{r})]+\frac{1}{2}\int d\mathbf{r}d\mathbf{r%
}^{\prime }U(\mathbf{r-r}^{\prime })G_{2}(\mathbf{r},\mathbf{r}^{\prime })
\notag \\
&&-\frac{1}{4M}\int d\mathbf{r}d\mathbf{r}^{\prime }G_{2}(\mathbf{r},\mathbf{%
r}^{\prime })[\nabla ^{2}\ln J(\mathbf{r},\mathbf{r}^{\prime })+\nabla \ln J(%
\mathbf{r},\mathbf{r}^{\prime })\nabla \ln n(\mathbf{r})],  \label{E2}
\end{eqnarray}%
where $\bar{g}_{2}(\mathbf{r},\mathbf{r}^{\prime })=G_{2}(\mathbf{r},\mathbf{%
r}^{\prime })/[n(\mathbf{r})n(\mathbf{r}^{\prime })J^{2}(\mathbf{r},\mathbf{r%
}^{\prime })]$.

We introduce the renormalized interaction%
\begin{equation}
U_{\mathrm{re}}(\mathbf{r,r}^{\prime })=\{U(\mathbf{r-r}^{\prime })J^{2}(%
\mathbf{r},\mathbf{r}^{\prime })+\frac{1}{M}[\nabla J(\mathbf{r},\mathbf{r}%
^{\prime })]^{2}+\frac{1}{4M}\nabla f(\mathbf{r},\mathbf{r}^{\prime })\nabla
\}\bar{g}_{2}(\mathbf{r},\mathbf{r}^{\prime })
\end{equation}%
to re-express the energy as%
\begin{equation}
E=\int d\mathbf{r[}\frac{1}{2M}\nabla \phi (\mathbf{r})\nabla \phi (\mathbf{r%
})+V(\mathbf{r})n(\mathbf{r})]+\frac{1}{2}\int d\mathbf{r}d\mathbf{r}%
^{\prime }U_{\mathrm{re}}(\mathbf{r,r}^{\prime })n(\mathbf{r})n(\mathbf{r}%
^{\prime }).
\end{equation}

\subsection{Variational energy for the state $\left\vert \Psi
_{N}\right\rangle $}

For the second ansatz $\left\vert \Psi _{N}\right\rangle $, the particle
number is conserved. The normalization factor, the density, and the second
order correlation function are%
\begin{equation}
\mathcal{N}=N!\int D[\mathbf{r}]\prod_{i<j(=1)}^{N}J^{2}(\mathbf{r}_{i},%
\mathbf{r}_{j})\prod_{j=1}^{N}\phi _{0}^{2}(\mathbf{r}_{j}),
\end{equation}%
\begin{equation}
n(\mathbf{r})=\int D[\mathbf{r}]\sum_{i}\delta (\mathbf{r}-\mathbf{r}%
_{i})P(\{\mathbf{r}\}),  \label{nr}
\end{equation}%
and%
\begin{equation}
G_{2}(\mathbf{r},\mathbf{r}^{\prime })=\int D[\mathbf{r}]\sum_{i\neq
j}\delta (\mathbf{r}-\mathbf{r}_{i})\delta (\mathbf{r}^{\prime }-\mathbf{r}%
_{j})P(\{\mathbf{r}\}),
\end{equation}%
where%
\begin{equation}
P(\{\mathbf{r}\})=\frac{\prod_{i<j=1}^{N}J^{2}(\mathbf{r}_{i},\mathbf{r}%
_{j})\prod_{j=1}^{N}\phi _{0}^{2}(\mathbf{r}_{j})}{\int D[\mathbf{r}%
]\prod_{i<j=1}^{N}J^{2}(\mathbf{r}_{i},\mathbf{r}_{j})\prod_{j=1}^{N}\phi
_{0}^{2}(\mathbf{r}_{j})}
\end{equation}%
is the probability distribution.

The ground state energy reads%
\begin{eqnarray}
E &=&\int d\mathbf{r}\frac{1}{2M}N\int D_{N-1}[\mathbf{r}]\{\nabla \lbrack
\prod_{j=1}^{N-1}J(\mathbf{r},\mathbf{r}_{j})\phi _{0}(\mathbf{r})]\}^{2}%
\frac{\prod_{i<j=1}^{N-1}J^{2}(\mathbf{r}_{i},\mathbf{r}_{j})%
\prod_{j=1}^{N-1}\phi _{0}^{2}(\mathbf{r}_{j})}{\int D[\mathbf{r}%
]\prod_{i<j=1}^{N}J^{2}(\mathbf{r}_{i},\mathbf{r}_{j})\prod_{j=1}^{N}\phi
_{0}^{2}(\mathbf{r}_{j})}  \notag \\
&&+\int d\mathbf{r}V(\mathbf{r})n(\mathbf{r})+\frac{1}{2}\int d\mathbf{r}d%
\mathbf{r}^{\prime }U(\mathbf{r-r}^{\prime })G_{2}(\mathbf{r},\mathbf{r}%
^{\prime })  \notag \\
&=&\int d\mathbf{r}\frac{1}{4M}N\int D_{N-1}[\mathbf{r}][\sum_{j=1}^{N-1}%
\nabla \ln J(\mathbf{r},\mathbf{r}_{j})+\nabla \ln \phi _{0}(\mathbf{r}%
)]\nabla \frac{\prod_{i<j=1}^{N}J^{2}(\mathbf{r}_{i},\mathbf{r}%
_{j})\prod_{j=1}^{N}\phi _{0}^{2}(\mathbf{r}_{j})}{\int D[\mathbf{r}%
]\prod_{i<j=1}^{N}J^{2}(\mathbf{r}_{i},\mathbf{r}_{j})\prod_{j=1}^{N}\phi
_{0}^{2}(\mathbf{r}_{j})}  \notag \\
&&+\int d\mathbf{r}V(\mathbf{r})n(\mathbf{r})+\frac{1}{2}\int d\mathbf{r}d%
\mathbf{r}^{\prime }U(\mathbf{r-r}^{\prime })G_{2}(\mathbf{r},\mathbf{r}%
^{\prime })  \notag \\
&=&-\int d\mathbf{r}\frac{1}{4M}N\int D_{N-1}[\mathbf{r}][\nabla ^{2}\ln
\phi _{0}(\mathbf{r})]\frac{\prod_{i<j=1}^{N}J^{2}(\mathbf{r}_{i},\mathbf{r}%
_{j})\prod_{j=1}^{N}\phi _{0}^{2}(\mathbf{r}_{j})}{\int D[\mathbf{r}%
]\prod_{i<j=1}^{N}J^{2}(\mathbf{r}_{i},\mathbf{r}_{j})\prod_{j=1}^{N}\phi
_{0}^{2}(\mathbf{r}_{j})}  \notag \\
&&-\int d\mathbf{r}d\mathbf{r}^{\prime }\frac{1}{4M}\nabla ^{2}\ln J(\mathbf{%
r},\mathbf{r}^{\prime })G_{2}(\mathbf{r},\mathbf{r}^{\prime })+\int d\mathbf{%
x}V(\mathbf{r})n(\mathbf{r})+\frac{1}{2}\int d\mathbf{r}d\mathbf{r}^{\prime
}U(\mathbf{r-r}^{\prime })G_{2}(\mathbf{r},\mathbf{r}^{\prime }),
\end{eqnarray}%
where the measure $D_{N-1}[\mathbf{r}]=d\mathbf{r}_{1}...d\mathbf{r}_{N-1}$.
Substituting $\phi _{0}(\mathbf{r})$ with $\phi (\mathbf{r})=\sqrt{n(\mathbf{%
r})}$ via Eq. (\ref{nr}), we obtain%
\begin{eqnarray}
E &=&\frac{1}{2M}\int d\mathbf{r}\nabla \phi (\mathbf{r})\nabla \phi (%
\mathbf{r})-\int d\mathbf{r}d\mathbf{r}^{\prime }\frac{1}{4M}\nabla \ln J(%
\mathbf{r},\mathbf{r}^{\prime })\nabla \ln n(\mathbf{r})G_{2}(\mathbf{r},%
\mathbf{r}^{\prime })  \notag \\
&&-\int d\mathbf{r}d\mathbf{r}^{\prime }\frac{1}{4M}\nabla ^{2}\ln J(\mathbf{%
r},\mathbf{r}^{\prime })G_{2}(\mathbf{r},\mathbf{r}^{\prime })+\int d\mathbf{%
x}V(\mathbf{r})n(\mathbf{r})+\frac{1}{2}\int d\mathbf{r}d\mathbf{r}^{\prime
}U(\mathbf{r-r}^{\prime })G_{2}(\mathbf{r},\mathbf{r}^{\prime }),
\end{eqnarray}%
which is exactly Eq. (\ref{E2}) derived from the first ansatz.

In general, the energy $E$ can be evaluated using the Monte Carlo sampling.
Alternatively, in the low and intermediate density regimes, the chain
diagram \cite{cluster1958} is taken into account in the resummation in Eq. (%
\ref{G2}), which results in the analytical expression%
\begin{equation}
G_{2}(\mathbf{r},\mathbf{r}^{\prime })=n(\mathbf{r})n(\mathbf{r}^{\prime
})J^{2}(\mathbf{r},\mathbf{r}^{\prime })[1+\int d\mathbf{r}_{1}f(\mathbf{r},%
\mathbf{r}_{1})n(\mathbf{r}_{1})F(\mathbf{r}_{1},\mathbf{r}^{\prime })].
\end{equation}%
Under the condition that the absolute value of the eigenvalues of $f(\mathbf{r},\mathbf{r}^{\prime
})\phi (\mathbf{r})\phi (\mathbf{r}^{\prime })$ are smaller than $1$, the
Dyson series%
\begin{equation}
F(\mathbf{r},\mathbf{r}^{\prime })=f(\mathbf{r},\mathbf{r}^{\prime })+\int d%
\mathbf{r}_{1}f(\mathbf{r},\mathbf{r}_{1})n(\mathbf{r}_{1})F(\mathbf{r}_{1},%
\mathbf{r}^{\prime })
\end{equation}%
is convergent.

The minimization of the functional $E[\phi (\mathbf{r}),J(\mathbf{r},\mathbf{%
r}^{\prime })]$ leads to the ground state configuration $\{\phi (\mathbf{r}%
),J(\mathbf{r},\mathbf{r}^{\prime })\}$, which determines the first order
correlation function%
\begin{eqnarray}
G_{1}(\mathbf{r},\mathbf{r}^{\prime }) &=&\phi (\mathbf{r})\phi (\mathbf{r}%
^{\prime })\exp \{\int d\mathbf{r}_{1}[J(\mathbf{r},\mathbf{r}_{1})J(\mathbf{r}%
^{\prime },\mathbf{r}_{1})-1]n(\mathbf{r}_{1})-\frac{1}{2}\int d\mathbf{r}%
_{1}[f(\mathbf{r},\mathbf{r}_{1})+f(\mathbf{r}^{\prime },\mathbf{r}_{1})]n(%
\mathbf{r}_{1})  \notag \\
&&+\frac{1}{2}\int d\mathbf{r}_{1}d\mathbf{r}_{2}F(\mathbf{r}_{1},\mathbf{r}%
_{2})\prod_{j=1,2}[J(\mathbf{r},\mathbf{r}_{j})J(\mathbf{r}^{\prime },%
\mathbf{r}_{j})-1]\prod_{j=1,2}n(\mathbf{r}_{j})  \notag \\
&&-\frac{1}{4}\int d\mathbf{r}_{1}d\mathbf{r}_{2}f(\mathbf{r},\mathbf{r}%
_{1})n(\mathbf{r}_{1})F(\mathbf{r}_{1},\mathbf{r}_{2})n(\mathbf{r}_{2})f(%
\mathbf{r}_{2},\mathbf{r})  \notag \\
&&-\frac{1}{4}\int d\mathbf{r}_{1}d\mathbf{r}_{2}f(\mathbf{r}^{\prime },%
\mathbf{r}_{1})n(\mathbf{r}_{1})F(\mathbf{r}_{1},\mathbf{r}_{2})n(\mathbf{r}%
_{2})f(\mathbf{r}_{2},\mathbf{r}^{\prime })\}
\end{eqnarray}%
via the chain diagram resummation in Eq. (\ref{G1}).

\section{Interactions, condensate fractions, and momentum distributions in
quasi-2D \label{sm:U2D}}

In this section, we derive the two-body interaction, the first order
correlation function, the condensate fraction, and the momentum distribution
in quasi-2D. Using the Gaussian wavefunction $\phi
_{0}(z)=e^{-z^{2}/(2a_{z}^{2})}/(\pi ^{1/4}a_{z}^{1/2})$, we obtain%
\begin{equation}
U_{2D}(\rho )=\frac{1}{\sqrt{2\pi }a_{z}}\int dzU(\mathbf{r})e^{-\frac{z^{2}%
}{2a_{z}^{2}}}.  \label{U2D}
\end{equation}%
The interaction in 3D reads%
\begin{equation}
U(\mathbf{r})=C_{3}\frac{2z^{2}-\rho ^{2}}{(\rho ^{2}+z^{2})^{5/2}}+C_{6}%
\frac{\rho ^{2}(\rho ^{2}+2z^{2})}{(\rho ^{2}+z^{2})^{5}}
\end{equation}%
in terms of $\rho $ and $z$.

Inserting $U(\mathbf{r})$ to Eq. (\ref{U2D}), we obtain the isotropic
interaction%
\begin{equation}
U_{2D}(\rho )=\frac{C_{3}}{a_{z}^{3}}u_{3}(\bar{\rho})+\frac{C_{6}}{a_{z}^{6}%
}u_{6}(\bar{\rho}),
\end{equation}%
where $\bar{\rho}=\rho /a_{z}$, and%
\begin{eqnarray}
u_{3}(\bar{\rho}) &=&\frac{1}{\sqrt{2\pi }}\int dt\frac{2t^{2}-\bar{\rho}^{2}%
}{(\bar{\rho}^{2}+t^{2})^{5/2}}e^{-\frac{t^{2}}{2}},  \notag \\
u_{6}(\bar{\rho}) &=&\frac{1}{\sqrt{2\pi }}\bar{\rho}^{2}\int dt\frac{\bar{%
\rho}^{2}+2t^{2}}{(\bar{\rho}^{2}+t^{2})^{5}}e^{-\frac{t^{2}}{2}}.
\end{eqnarray}%
The integrals can be evaluated analytically as%
\begin{eqnarray}
u_{3}(\bar{\rho}) &=&\frac{1}{2\sqrt{2\pi }}e^{\frac{\bar{\rho}^{2}}{4}}[%
\bar{\rho}^{2}K_{1}(\frac{\bar{\rho}^{2}}{4})-(\bar{\rho}^{2}+2)K_{0}(\frac{%
\bar{\rho}^{2}}{4})],  \notag \\
u_{6}(\bar{\rho}) &=&\frac{1}{384\sqrt{2\pi }\bar{\rho}^{5}}[\sqrt{2\pi }%
\bar{\rho}(\bar{\rho}^{6}+11\bar{\rho}^{4}-39\bar{\rho}^{2}+135)  \notag \\
&&+\pi e^{\frac{\bar{\rho}^{2}}{2}}(\bar{\rho}^{8}+12\bar{\rho}^{6}-30\bar{%
\rho}^{4}+84\bar{\rho}^{2}-135)(\text{erf}(\frac{\bar{\rho}}{\sqrt{2}})-1)],
\end{eqnarray}%
where $K_{0,1}(x)$ and erf$(x)$ are Bessel and error functions.

Due to the rotational symmetry in 2D, $G_{1}({\boldsymbol{\rho }},{%
\boldsymbol{\rho }}^{\prime })=\sum_{m}G_{1m}(\rho ,\rho ^{\prime
})e^{im(\varphi -\varphi ^{\prime })}/(2\pi )$ can be calculated
numerically. By diagonalizing $G_{1m}(\rho ,\rho ^{\prime })$ in different
angular momentum channels $m$, we find the largest eigenvalue $N_{0}$ in the
channel $m=0$ and the corresponding eigenstate $\bar{\varphi}_{0}(\rho )$
characterizing the condensate wavefunction, where the condensate fraction $%
f_{c}=N_{0}/N$.

The momentum distribution%
\begin{equation}
\tilde{n}(\mathbf{k})=\left\langle c_{\mathbf{k}}^{\dagger }c_{\mathbf{k}%
}\right\rangle =\int \frac{d^{2}{\boldsymbol{\rho }}d^{2}{\boldsymbol{\rho }}%
^{\prime }}{(2\pi )^{2}}e^{i\mathbf{k}\cdot {\boldsymbol{\rho }}^{\prime }-i%
\mathbf{k}\cdot {\boldsymbol{\rho }}}G_{1}({\boldsymbol{\rho }},{\boldsymbol{%
\rho }}^{\prime })
\end{equation}%
is the Fourier transform of $G_{1}({\boldsymbol{\rho }},{\boldsymbol{\rho }}%
^{\prime })$. In terms of $G_{1m}(\rho ,\rho ^{\prime })$, we can calculate%
\begin{equation}
\tilde{n}(\mathbf{k})=\frac{1}{2\pi }\sum_{m}\int \rho d\rho \rho ^{\prime
}d\rho ^{\prime }J_{m}(k\rho )G_{1m}(\rho ,\rho ^{\prime })J_{m}(k\rho
^{\prime })
\end{equation}%
numerically.

\end{document}